\documentstyle[12pt]{article}
\begin{document}
\centerline{\Large\bf Einstein-Maxwell and Einstein-Proca theory}
\vskip 0.1in
\centerline{\Large\bf from a modified gravitational action}
\vskip .7in
\centerline{Dan N. Vollick}
\centerline{Irving K. Barber School of Arts and Sciences}
\centerline{University of British Columbia Okanagan}
\centerline{3333 University Way}
\centerline{Kelowna, B.C.}
\centerline{V1V 1V7}
\vskip .9in
\centerline{\bf\large Abstract}
\vskip 0.5in
\noindent
A modified gravitational action is considered which involves the quantity
$F_{\mu\nu}=\partial_{\mu}\Gamma_{\nu}-\partial_{\nu}\Gamma_{\mu}$, where
$\Gamma_{\mu}=\Gamma^{\alpha}_{\mu\alpha}$. Since $\Gamma_{\mu}$ transforms
like a $U(1)$ gauge field under coordinate transformations terms such as $F^{\mu\nu}F_{\mu\nu}$ are invariant under coordinate transformations. If
such a term is added to the usual gravitational action the resulting 
field equations, obtained from a Palatini variation, are the Einstein-Proca
equations. The vector field can be coupled to point charges or to a complex
scalar density of weight $ie$, where $e$ is the charge of the field. If this
scalar density is taken to be $g^{-ie/2}$ and the overall factor of the scalar
density Lagrangian takes on a particular value the resulting field equations
are the Einstein-Maxwell equations.   
\newpage
\section*{Introduction}
In the standard formulation of the general theory of relativity non-gravitational
fields are not related to the geometry of space-time, but are fields that
exist in space-time. In this paper I will show that both 
massive and massless geometrical vector fields can be obtained in the Palatini formalism from a modified gravitational action.

The contraction of the connection $\Gamma_{\mu}=\Gamma^{\alpha}_{\mu\alpha}$
is not a vector but has the following transformation law
\begin{equation}
\bar{\Gamma}_{\mu}=\frac{\partial x^{\nu}}{\partial\bar{x}^{\mu}}\Gamma_{\nu}
+\frac{\partial}{\partial \bar{x}^{\mu}}
\left|\frac{\partial x}{\partial\bar{x}}\right|\; ,
\label{eq1}
\end{equation}
where $|\partial x/\partial\bar{x}|$ 
is the Jacobian of the transformation. It is interesting to compare
this transformation law to that of a $U(1)$ gauge field $A_{\mu}$ 
under a combined gauge and coordinate transformation. Under a gauge
transformation 
\begin{equation}
A^{'}_{\mu}=A_{\mu}+\frac{1}{e}\frac{\partial\chi}{\partial x^{\mu}}\; ,
\end{equation}
where $\chi$ is an arbitrary function and $e$ is the charge of the 
field. Performing a coordinate transformation gives
\begin{equation}
\bar{A}_{\mu}=\frac{\partial x^{\nu}}{\partial\bar{x}^{\mu}}A_{\nu}+
\frac{1}{e}\frac{\partial\chi}
{\partial\bar{x}^{\mu}}\; .
\end{equation}
This is the same as (\ref{eq1}) if we take $\chi=\frac{1}{e}ln |\partial
x/\partial\bar{x}|$. Thus, under a coordinate
transformation the field $\Gamma_{\mu}$ behaves like a $U(1)$ gauge field
under a combined gauge and coordinate transformation.
This implies that the quantity 
$F_{\mu\nu}=\partial_{\mu}\Gamma_{\nu}-\partial_{\nu}\Gamma_{\mu}$ is a tensor
under general coordinate transformations.
In general relativity $\Gamma_{\mu}=\partial\; ln(\sqrt{g})/\partial x^{\mu}$ and the associated field strength $F_{\mu\nu}=\partial_{\mu}
\Gamma_{\nu}-\partial_{\nu}\Gamma_{\mu}$ vanishes because $\Gamma_{\mu}$ is
a pure gauge. It is possible however, that in modified theories this is not
the case. 
In fact, I have shown {\cite{Vo1} that $\Gamma_{\mu}$ is not necessarily
a pure gauge in Born-Infeld-Einstein theory. 
In this paper I consider adding a term of the form $F^{\mu\nu}F_{\mu\nu}$
to the standard gravitational action and show that, under a Palatini 
variation 
\begin{equation}
\Gamma_{\mu}=\frac{\partial \;ln(\sqrt{g})}{\partial x^{\mu}}+V_{\mu}\; ,
\end{equation}
where $V_{\mu}$ is an arbitrary vector. The field equation satisfied by
$V_{\mu}$ is shown to be the Proca equation and the gravitational field equations
are the Einstein-Proca equations.

Two types of sources for the vector field are considered. First it is shown that
point charges can be coupled to the field. Then, following the formalism
for Abelian gauge fields, it is shown that a complex scalar density of weight
$ie$ can be coupled to the vector field. Here $e$ is the charge of the field
that appears in the covariant derivative. There is a geometric quantity,
namely $g^{-ie/2}$, that is a scalar density of weight $ie$. The Lagrangian
for this scalar density is
\begin{equation}
L=-\frac{1}{4}\beta\sqrt{g}g^{-2}\nabla_{\mu}g\nabla^{\mu}g\; ,
\end{equation}
where $\beta$ is a constant. It is shown that if $\beta=-3/(4\kappa)$ the vector
field is massless and we obtain the Einstein-Maxwell equations.
\section*{The Field Equations}
The field equations follow from the variation of the action
\begin{equation}
S=\int\left[ -\frac{1}{2\kappa}R-\frac{\alpha}{4}F^{\mu\nu}F_{\mu\nu}+L_M\right]\sqrt{g}
\; d^4x\;
,
\label{action}
\end{equation}
where
\begin{equation}
R_{\mu\nu}=\partial_{\nu}\Gamma^{\alpha}_{\mu\alpha}-
\partial_{\alpha}\Gamma^{\alpha}_{\mu\nu}+\Gamma^{\alpha}_{\beta\mu}
\Gamma^{\beta}_{\alpha\nu}-\Gamma^{\beta}_{\mu\nu}\Gamma^{\alpha}_
{\beta\alpha}\; ,
\end{equation}
\begin{equation}
F_{\mu\nu}=\partial_{\mu}\Gamma_{\nu}-\partial_{\nu}\Gamma_{\mu}\;,
\end{equation}
$R=g^{\mu\nu}R_{\mu\nu}$, 
$\Gamma_{\mu}=\Gamma^{\alpha}_{\mu\alpha}$, $\kappa=8\pi G$,
$L_M$ is the matter Lagrangian and the connection is taken to be symmetric.
Here I will use a Palatini variation of the
action, which treats $g_{\mu\nu}$ and $\Gamma^{\alpha}_{\mu\nu}$ as independent
variables. 
Varying the action with respect to $g_{\mu\nu}$ gives
\begin{equation}
G_{(\mu\nu)}(\Gamma)=-\alpha\kappa \left[ F_{\mu\alpha}F^{\;\;\alpha}_{\nu}-\frac{1}{4}
g_{\mu\nu}F^{\alpha\beta}F_{\alpha\beta}\right]-\kappa T_{\mu\nu}\; ,
\label{Einstein1}
\end{equation}
where $G_{(\mu\nu)}(\Gamma)$ is the symmetric part of the Einstein tensor, 
which depends on $\Gamma^{\alpha}_{\mu\nu}$,
and $T_{\mu\nu}$ is the energy-momentum tensor of the matter.

Varying the action with respect to $\Gamma^{\alpha}_{\mu\nu}$ gives
\begin{equation}
\nabla_{\alpha}(\sqrt{g}g^{\mu\nu})-\frac{1}{2}\left\{\left[\nabla_{\beta}(\sqrt{g}g^{\beta\mu})
+2\alpha\kappa\sqrt{g}\;\tilde{\nabla}_{\beta}F^{\beta\mu}\right]\delta^{\nu}_{\alpha}
+\left[\nabla_{\beta}(\sqrt{g}g^{\beta\nu})
+2\alpha\kappa\sqrt{g}\;\tilde{\nabla}_{\beta}F^{\beta\nu}\right]\delta^{\mu}_{\alpha}
\right\}=0\; ,
\label{gamma}
\end{equation}
where $\tilde{\nabla}$ is the covariant derivative with respect to the Christoffel
symbol. Contracting over $\mu$ and $\alpha$ gives
\begin{equation}
\nabla_{\beta}(\sqrt{g}g^{\beta\mu})=-\frac{10}{3}\alpha\kappa\sqrt{g}\;\tilde{\nabla}
_{\beta}F^{\beta\mu}
\label{contract}
\end{equation}
and substituting this into (\ref{gamma}) gives 
\begin{equation}
\nabla_{\alpha}(\sqrt{g}g^{\mu\nu})-\frac{1}{5}\left[\delta^{\nu}_{\alpha}
\nabla_{\beta}(\sqrt{g}g^{\beta\mu})+\delta^{\mu}_{\alpha}
\nabla_{\beta}(\sqrt{g}g^{\beta\nu})\right]=0\;.
\label{gamma1}
\end{equation}
Since the trace of the left hand side of (\ref{gamma1}) vanishes there are four
too few equations and the system is under determined. Thus, we expect four
arbitrary functions in the solution. Such a solution is given by
\begin{equation}
\nabla_{\alpha}\left[\sqrt{g}g^{\mu\nu}\right]=
-\sqrt{g}\left[\delta^{\mu}_{\alpha}V^{\nu}+\delta^{\nu}_{\alpha}
V^{\mu}\right]\;,
\label{vector}
\end{equation}
where $V^{\mu}$ is an arbitrary vector. The connection that follows from
this set of equations is given by \cite{Vo1}
\begin{equation}
\Gamma^{\alpha}_{\mu\nu}=\left\{
\begin{array}{ll}
\alpha\\
\mu\nu\\
\end{array}
\right\}-\frac{1}{2}\left[3g_{\mu\nu}V^{\alpha}-\delta^{\alpha}_{\mu}V_{\nu}-
\delta^{\alpha}_{\nu}V_{\mu}\right]\; ,
\label{connection}
\end{equation}
where the first term on the right hand side is the Christoffel symbol. Thus,
\begin{equation}
\Gamma_{\mu}=\frac{\partial\; ln(\sqrt{g})}{\partial x^{\mu}}+V_{\mu}
\end{equation}
and $F_{\mu\nu}=\tilde{\nabla}_{\mu}V_{\nu}-\tilde{\nabla}_{\nu}V_{\mu}$.

Substituting (\ref{vector}) into (\ref{contract}) gives
\begin{equation}
\tilde{\nabla}_{\beta}F^{\beta\mu}=\frac{3}{2\alpha\kappa}V^{\mu}\; ,
\label{em}
\end{equation}
so that $V_{\mu}$ satisfies the Proca equation and has mass $\sqrt{3/(2\alpha
\kappa)}$. This implies that $\Gamma_{\mu}$ satisfies the field equation
\begin{equation}
\tilde{\nabla}_{\beta}F^{\beta\mu}=\frac{3}{2\alpha\kappa}\left(\Gamma^{\mu}-
\frac{\partial\; ln(\sqrt{g})}{\partial x^{\mu}}\right)\; ,
\end{equation}
where $F_{\mu\nu}=\partial_{\mu}\Gamma_{\nu}-\partial_{\nu}\Gamma_{\mu}$
is a tensor. Thus, $\Gamma_{\mu}$ satisfies the Proca equation with a source.
Note that even though $\Gamma_{\mu}$ is a massive field the field equations
are invariant under the transformation (\ref{eq1}) since the source transforms
in a similar way.

To see that (\ref{Einstein1}) are the gravitational field equations associated
with the Proca field it is necessary to express them in terms of the
Einstein tensor $\tilde{G}_{\mu\nu}$, which is defined
in terms of the Christoffel symbol. The relationship between the Ricci tensors
is given by
\begin{equation}
R_{\mu\nu}(\Gamma)=\tilde{R}_{\mu\nu}-\tilde{\nabla}_{\alpha}H^{\alpha}_{\mu\nu}+\tilde{\nabla}
_{\nu}H^{\alpha}_{\alpha\mu}-H^{\alpha}_{\alpha\beta}H^{\beta}_{\mu\nu}+H^{\alpha}
_{\mu\beta}H^{\beta}_{\alpha\nu}\; ,
\end{equation}
where $H^{\alpha}_{\mu\nu}$ is the tensor field
\begin{equation}
H^{\alpha}_{\mu\nu}=\Gamma^{\alpha}_{\mu\nu}-\left\{
\begin{array}{ll}
\alpha\\
\mu\nu\\
\end{array}
\right\}
\end{equation}
and $\tilde{\nabla}$ is the metric compatible covariant derivative.
A simple calculation shows that
\begin{equation}
G_{(\mu\nu)}(\Gamma)=\tilde{G}_{\mu\nu}+\frac{3}{2}\left[V_{\mu}V_{\nu}-\frac{1}{2}
g_{\mu\nu}\left(2\tilde{\nabla}_{\alpha}V^{\alpha}+V^{\alpha}V_{\alpha}\right)
\right]\; .
\label{G}
\end{equation}
From (\ref{em}) it is easy to see that $\tilde{\nabla}_{\alpha}V^{\alpha}=0$.
Thus, the Einstein field equations are given by
\begin{equation}
\tilde{G}_{\mu\nu}=-\alpha\kappa \left[F_{\mu\alpha}F^{\;\;\alpha}_{\nu}-\frac{1}{4}
g_{\mu\nu}F^{\alpha\beta}F_{\alpha\beta}+\frac{3}{2\alpha\kappa}(V_{\mu}V_{\nu}
-\frac{1}{2}g_{\mu\nu}V^{\alpha}V_{\alpha})\right]-\kappa T_{\mu\nu}
\label{Einstein2}
\end{equation}
and are the field equations for a Proca field plus any additional matter
that contributes to $T_{\mu\nu}$. Note that if $\alpha\sim 1$ the mass of
the vector field is of order of the Planck mass.
  
It is interesting to note that the transformation law of the full connection
\begin{equation}
\bar{\Gamma}^{\mu}_{\alpha\beta}=\frac{\partial\bar{x}^{\mu}}{\partial x^{\lambda}}
\frac{\partial x^{\sigma}}{\partial\bar{x}^{\alpha}}\frac{\partial x^{\kappa}}
{\partial\bar{x}^{\beta}}\Gamma^{\lambda}_{\sigma\kappa}+\frac{\partial}
{\partial\bar{x}^{\alpha}}\left[\frac{\partial x^{\sigma}}{\partial\bar{x}
^{\beta}}\right]\frac{\partial\bar{x}^{\mu}}{\partial x^{\sigma}}
\end{equation}
is identical with the transformation law for the Yang-Mills potential
\begin{equation}
(\bar{A}_{\mu})^i_{\;\; k}=\frac{\partial x^{\nu}}{\partial\bar{x}^{\mu}}
S^i_{\;\; l}(A_{\nu})^l_{\;\; m}(S^{-1})^m_{\;\; k}
-\frac{i}{e}\frac{\partial S^i_{\; m}}{\partial\bar{x}_{\mu}}
(S^{-1})^m_{\;\; k}\; .
\end{equation}
if we take
\begin{equation}
S^{i}_{\;\; k}=\frac{\partial x^{i}}{\partial\bar{x}^{k}}
\end{equation}
and 
\begin{equation}
\Gamma^l_{\alpha k}=ie(A_{\alpha})^l_{\;\; k}
\end{equation}
Note that both Latin and Greek indices are space-time indices and run from
0 to 3. The field strength tensor is proportional to the Riemann tensor,
but the action is taken to be proportional to $R$, not to the Riemann tensor
squared (see \cite{De1} for a more detailed discussion on the Yang-Mills
fields associated with diffeomorphisms). Thus, we can think of general relativity
as the gauge theory associated with diffeomorphisms (see \cite{He1} for alternative
approaches to gauge theories of gravity).

\section*{Sources of the vector field}
I will first consider coupling point particles to the vector field.
The simplest
coupling is given by
\begin{equation}
L_c=-\alpha\sqrt{g}\;\Gamma_{\mu}
J^{\mu}\; 
\label{L2}
\end{equation}
where $J^{\mu}$ is the conserved current associated with the source. For
a point particle with ``charge" $e$
\begin{equation}
J^{\mu}(x^{\alpha})=\frac{e}{\sqrt{g}}\int \frac{dx^{\mu}(p)}{dp}
\delta(x^{\alpha}
-x^{\alpha}(p))dp
\label{J}
\end{equation}
where $p$ is a parameterization of the particle's
world line. Of course, the particles that produce
$J^{\mu}$ must also appear in $L_M$.

At first sight it may appear that there is a problem with this 
Lagrangian since
$\Gamma_{\mu}$ is not a vector. An analogous situation occurs
in electrodynamics where the interaction Lagrangian $\sqrt{g}A_{\mu}J^{\mu}$
appears not to be gauge invariant. However, if $\tilde{\nabla}_{\mu}J^{\mu}
=0$ the Lagrangian only changes by a total derivative under a gauge
transformation. Now, under a coordinate transformation
\begin{equation}
\bar{\Gamma}_{\mu}=\frac{\partial x^{\nu}}
{\partial \bar{x}^{\mu}}\Gamma_{\nu}+\frac{\partial}{\partial
\bar{x}^{\mu}}\ln\left|\frac{\partial x}{\partial\bar{x}}\right|
\label{trans}
\end{equation}
where $|\partial x/\partial\bar{x}|$ is the Jacobian of the transformation.
This is analogous to a gauge transformation and it is easy to see
that the Lagrangian $L_c$ only changes by a total derivative
if $\tilde{\nabla}_{\mu}J^{\mu}=0$. 

Varying the action with respect to $g_{\mu\nu}$ gives (\ref{Einstein1}) and
varying the action with respect to $\Gamma^{\alpha}_{\mu\nu}$ gives (\ref{gamma})
with
\begin{equation}
\tilde{\nabla}_{\beta}F^{\beta\mu}\rightarrow\tilde{\nabla}_{\beta}F^{\beta\mu}
-J^{\mu}\; .
\end{equation}
One interesting
property of this choice of Lagrangian is that the current $J^{\mu}$
does not enter into the equation defining the connection, so 
that (\ref{connection}) is still valid. The field $F^{\mu\nu}$ now
satisfies the equation
\begin{equation}
\tilde{\nabla}_{\beta}F^{\beta\mu}=\frac{3}{2\alpha\kappa}V^{\mu}
+J^{\mu}
\end{equation}
which is the Proca equation with source $J^{\mu}$. This equation tells us
that $\tilde{\nabla}_{\mu}V^{\mu}=0$ and that (\ref{Einstein2}) holds.
The field equations therefore correspond to a Proca equation with source minimally coupled to gravity.

Next consider coupling a complex scalar density to the vector 
field. In a $U(1)$
gauge theory consisting of a complex scalar field $\phi$ coupled to a 
gauge field $A_{\mu}$ the fields transform as
\begin{equation}
\bar{\phi}=e^{-i\chi}\phi\; ,
\label{phi}
\end{equation}
\begin{equation}
\bar{A}_{\mu}=\frac{\partial x^{\nu}}{\partial\bar{x}^{\mu}}A_{\nu}+
\frac{1}{e}\frac{\partial\chi}{\partial\bar{x}^{\mu}}; ,
\label{A}
\end{equation}
under a combined gauge and coordinate transformation,
the covariant derivative of $\phi$ is given by
\begin{equation}
D_{\mu}\phi=(\partial_{\mu}+ieA_{\mu})\phi\; ,
\end{equation}
and the Lagrangian is given by
\begin{equation}
L=-\frac{1}{2}D_{\mu}\phi D^{\mu}\phi^*-\frac{1}{2}m^2\phi^*\phi
-\frac{1}{4}F^{\mu\nu}F_{\mu\nu}\;,
\label{Lag1}
\end{equation}
where
\begin{equation}
D_{\mu}\phi^*=\left(\partial_{\mu}-ieA_{\mu}\right)\phi^*\; .
\end{equation}
In the theory presented here $A_{\mu}$ will be replaced by $\Gamma_{\mu}$.
From (\ref{trans}) and (\ref{A}) we see that
\begin{equation}
\chi=e\ln\left|\frac{\partial x}{\partial\bar{x}}\right|
\end{equation}
and from (\ref{phi}) we see that
\begin{equation}
\bar{\phi}=\left|\frac{\partial x}{\partial\bar{x}}\right|^{-ie}\phi\;.
\end{equation}
Thus, $\phi$ is a scalar density of complex weight $ie$. With $D_{\mu}\phi$
defined as
\begin{equation}
D_{\mu}=\left(\partial_{\mu}+ie\Gamma_{\mu}\right)\phi
\end{equation}
it is easy to show that
\begin{equation}
\bar{D}_{\mu}\bar{\phi}=\left|\frac{\partial x}{\partial\bar{x}}\right|^{-ie}
\frac{\partial x^{\alpha}}{\partial\bar{x}^{\mu}}D_{\alpha}\phi\; .
\end{equation}
In fact, $D_{\mu}\phi$ is just the covariant derivative of the tensor density
$\phi$ with respect to the connection $\Gamma^{\alpha}_{\mu\nu}$ and I will
write $\nabla_{\mu}\phi$ instead of $D_{\mu}\phi$.
The Lagrangian (\ref{Lag1}) is therefore a scalar under coordinate transformations.

The action for the theory will therefore be given by (\ref{action}) with
$L_M$ given by the scalar density terms in (\ref{Lag1}).
Varying the action with respect to $\phi^*$ gives
\begin{equation}
\nabla_{\mu}\left[\sqrt{g}\;\nabla^{\mu}\phi\right]-m^2\phi=0\; .
\end{equation}
Using
\begin{equation}
\nabla_{\mu}\sqrt{g}=-\sqrt{g}\;V_{\mu}
\end{equation}
gives the field equation
\begin{equation}
\nabla_{\mu}\nabla^{\mu}\phi-V_{\mu}\nabla^{\mu}\phi-m^2\phi=0\; .
\label{phi}
\end{equation}
The field equations that follow from varying the action with 
respect to $g_{\mu\nu}$ are
\begin{equation}
G_{(\mu\nu)}=-\alpha\kappa \left[ F_{\mu\alpha}F^{\;\;\alpha}_{\nu}-\frac{1}{4}
g_{\mu\nu}F^{\alpha\beta}F_{\alpha\beta}\right]-\kappa\left[\nabla_{(\mu}
\phi \nabla_{\nu)}\phi^*-\frac{1}{2}g_{\mu\nu}\left(\nabla_{\alpha}\phi \nabla^{\alpha}\phi^*+m^2
\phi^*\phi\right)\right]\; .
\label{gravity}
\end{equation}
Varying the action with respect to $\Gamma^{\alpha}_{\mu\nu}$ gives (\ref{gamma})
with
\begin{equation}
\tilde{\nabla}_{\beta}F^{\beta\mu}\rightarrow\tilde{\nabla}_{\beta}F^{\beta\mu}
-J^{\mu}\; ,
\end{equation}
where
\begin{equation}
J^{\mu}=\frac{ie}{2\alpha}\left[\phi \nabla^{\mu}\phi^*-\phi^*\nabla^{\mu}\phi\right]\;.
\label{current}
\end{equation}
The field $F^{\mu\nu}$ satisfies the equation
\begin{equation}
\tilde{\nabla}_{\beta}F^{\beta\mu}=\frac{3}{2\alpha\kappa}V^{\mu}
+J^{\mu}\; ,
\label{em3}
\end{equation}
which is the Proca equation with source $J^{\mu}$.
From (\ref{phi}) and (\ref{current}) it is easy to see
that 
\begin{equation}
\nabla_{\mu}J^{\mu}=V_{\mu}J^{\mu}\; ,
\end{equation}
 which implies that
\begin{equation}
\tilde{\nabla}_{\mu}J^{\mu}=0\; .
\label{div}
\end{equation}
Equations (\ref{em3}) and (\ref{div}) give
\begin{equation}
\tilde{\nabla}_{\mu}V^{\mu}=0\; .
\end{equation}
This together with (\ref{G}) and (\ref{gravity}) gives
\begin{equation}
\begin{array}{ll}
\tilde{G}_{\mu\nu}=-\alpha\kappa \left[ F_{\mu\alpha}F_{\nu}^{\;\;\;\alpha}-\frac{1}{4}
g_{\mu\nu}F^{\alpha\beta}F_{\alpha\beta}+\frac{3}{2\alpha\kappa}(V_{\mu}V_{\nu}
-\frac{1}{2}g_{\mu\nu}V^{\alpha}V_{\alpha})\right]\\
     \\
\;\;\;\;\;\;\;\;\;\;\;\;\;-\kappa\left[\nabla_{(\mu}
\phi \nabla_{\nu)}\phi^*-\frac{1}{2}g_{\mu\nu}\left(\nabla_{\alpha}\phi \nabla^{\alpha}\phi^*+m^2
\phi^*\phi\right)\right]\; .
\end{array}
\end{equation}
It is interesting to note that
\begin{equation}
\nabla_{\mu}\phi=\left(\tilde{\nabla}_{\mu}+ieV_{\mu}\right)\phi\equiv\tilde{D}_{\mu}\phi\;,
\end{equation}
where $\tilde{D}_{\mu}$ is the standard covariant derivative operator with
respect to the Christoffel symbol and that equation (\ref{phi}) can be written
as
\begin{equation}
\tilde{D}_{\mu}\tilde{D}^{\mu}\phi-m^2\phi=0\; .
\end{equation}
Thus, the above set of equations represents Einstein gravity minimally 
coupled to a massive vector field $V_{\mu}$ and a complex scalar density field $\phi$.
\section*{A geometric scalar density and the Einstein-Maxwell equations}
A simple geometric quantity that is a scalar density of weight $ie$
is $g^{-ie/2}$. In this case the scalar density is not an independent
variable and the field equations will differ from the ones given in the previous
section. The scalar density Lagrangian can be obtained from 
(\ref{Lag1}) by letting $\phi\rightarrow g^{-ie/2}$ and is given by
\begin{equation}
L=-\frac{1}{4}\beta\sqrt{g}\left[g^{-2}\nabla_{\mu}g\nabla^{\mu}g\right]\; ,
\end{equation}
where I have introduced an overall constant $\beta$. I have also taken $m=0$
fr simplicity since it only corresponds to a cosmological constant. 
The variation with respect to the connection gives (\ref{connection})
as before and (\ref{em3}) with $J^{\mu}$ given by
\begin{equation}
J^{\mu}=-\frac{\beta}{\alpha g}\nabla^{\mu}g\; .
\end{equation}
Now from (\ref{vector}) one can show that 
\begin{equation}
\nabla_{\alpha}g=-2gV_{\alpha}\; .
\end{equation}
The vector field equation becomes
\begin{equation}
\tilde{\nabla}_{\beta}F^{\beta\mu}=\left[\frac{3}{2\alpha\kappa}+\frac{2\beta}{\alpha}
\right]V^{\mu}\; .
\end{equation}
Variation of the action with respect to $g_{\mu\nu}$ gives
\begin{equation}
\tilde{G}_{\mu\nu}=-\alpha\kappa \left[F_{\mu\alpha}F_{\;\;\;\nu}^{\alpha}-\frac{1}{4}
g_{\mu\nu}F^{\alpha\beta}F_{\alpha\beta}\right]-\left(2\kappa\beta+\frac{3}{2}\right)
\left[V_{\mu}V_{\nu}
-\frac{1}{2}g_{\mu\nu}V^{\alpha}V_{\alpha}
\right]\; .
\end{equation} 
Thus, if $\beta=-3/(4\kappa)$ the resulting equations are the Einstein-Maxwell
equations. Point charge sources for the electromagnetic field can be included
by adding the Lagrangian (\ref{L2}) with $J^{\mu}$ given by (\ref{J}). Of
course, the Lagrangian for the point sources also has to be included and
will contribute a source term to the Einstein Field equations.
\section*{Conclusion}
A term proportional to $F^{\mu\nu}F_{\mu\nu}$, where $F_{\mu\nu}=\partial_{\mu}
\Gamma_{\nu}-\partial_{\nu}\Gamma_{\mu}$ and $\Gamma_{\mu}=\Gamma^{\alpha}_{\mu\alpha}$,
was added to the standard gravitational Lagrangian. Since the field $\Gamma_{\mu}$
transforms like a $U(1)$ gauge field under coordinate transformations the
field strength $F_{\mu\nu}$ is a tensor. The field equations obtained from
a Palatini variation are the Einstein-Proca equations. It was shown that
the vector field could be coupled to point charges or to a complex scalar
density of weight $ie$. If the scalar density is taken to be $g^{-ie/2}$
and the over all factor of the scalar density Lagrangian takes on a
particular value the resulting equations were shown to be the Einstein-Maxwell equations.
\section*{Acknowledgements}
This work was supported by the Natural Sciences and Engineering Research
Council of Canada.

\end{document}